\documentclass{article}
\usepackage[T1]{fontenc}
\usepackage{lmodern}
\usepackage{silence}
\usepackage{amsmath,amsfonts}
\usepackage{amsthm}
\newtheorem{definition}{Definition}

\usepackage{algorithm}
\usepackage{algorithmic}

\usepackage{graphicx}
\usepackage{pgfplots}
\pgfplotsset{compat=1.18}
\usepackage{tikz}
\usepackage{transparent}

\usepackage{array}
\usepackage{booktabs}
\usepackage{multicol}
\usepackage{multirow}

\usepackage{anyfontsize}
\usepackage{balance}
\usepackage{enumitem}
\usepackage{stfloats}

\usepackage{cite}
\usepackage{textcomp}
\usepackage{url}
\usepackage{verbatim}

\usepackage{hyperref}

\definecolor{FirstPlace}{rgb}{0.2, 0.6, 0.2}
\definecolor{SecondPlace}{rgb}{0.5, 0.8, 0.2}
\definecolor{ThirdPlace}{rgb}{0.8, 0.8, 0.2}

\definecolor{Biggest}{rgb}{0.2, 0.6, 0.2}
\definecolor{Lowest}{RGB}{164, 22, 0}

\newcommand{\biggest}[1]{\textcolor{Biggest}{\textbf{#1}}}
\newcommand{\lowest}[1]{\textcolor{Lowest}{\textbf{#1}}}

\begin{document}
\date{}
\title{Graph Reduction in Multirelational Networks: \\A Spreading-Oriented Reduction Benchmark}

\author{Mateusz Stolarski,
Micha{\l} Czuba,
Piotr Bielak,
Piotr Br{\'o}dka \\
Wroc{\l}aw University of Science and Technology, Wroc{\l}aw, Poland
\thanks{Corresponding author: Mateusz Stolarski (email: mateusz.stolarski@pwr.edu.pl)}
\thanks{The source code will be released after the review process.}
}


\maketitle

\begin{abstract}
Real-world networks are inherently incomplete, noisy, and dynamically evolving, making it difficult to capture all actors and their relationships. Their scale often renders direct analysis computationally demanding. While influence maximisation (IM) has been widely studied, the role of graph reduction as a preprocessing step, and its impact on IM accuracy, remains underexplored. In this work, we introduce the \textit{Spreading-Oriented Reduction Benchmark} (\textit{SORB}), an open-source, standardised framework for systematically evaluating IM models across diverse task settings. SORB provides an extensible pipeline operating on a representative collection of real-world networks, including single- and multilayer structures, and accounts for graph reduction directly into the evaluation process. This design shifts the focus from analysing IM algorithms in isolation to quantifying how graph reduction alters predictive performance. Using \textit{SORB}, we study the effects of sparsification and coarsening across multiple IM scenarios. Our results show that the impact of reduction is strongly dependent on both the network type (single-layer vs. multirelational) and the downstream task ($Gain@k$ vs. $\mathrm{AUC}_{\mathrm{cutoff}}$): sparsification preserves seed set quality on single-layer networks, whereas flattened multilayer networks exhibit systematic ranking degradation regardless of reduction strategy. These findings highlight the importance of reduction-aware, multi-task evaluation when studying spreading processes in complex networks.
\end{abstract}


\section{Introduction}
Complex networks arise naturally in many real-world systems, including social interactions, biological processes, and technological infrastructures. They are typically represented by graphs constructed from relational data in order to capture different types of connections and the processes occurring between the actors that constitute such systems~\cite{easley2010networks}.

One of the most widely studied problems on complex network systems is the spreading phenomenon, which can be observed in various domains such as epidemiology or social media~\cite{magnani2015spreadingmln}. In such contexts, multilayer network models might be applied to provide a more faithful representation of the processes taking place. A closely related and extensively investigated task is the influence maximisation problem, which aims to identify a set of seed nodes that maximises the expected spread of information or influence in a network. This problem was formally introduced in~\cite{kempe2003maximizing}, which established the foundations for algorithmic approaches to modelling diffusion processes in social networks. However, tasks related to spreading processes and influence maximisation can take different forms depending on how the problem is defined~\cite{czuba2025identifying,kempe2003maximizing,singh2022influence}.

When dealing with network data, an important factor is its completeness, as this allows the characteristics of the analysed system to be reproduced as faithfully as possible. In real-world scenarios, however, which are characterised by dynamically evolving phenomena, achieving a complete mapping of all network actors and the relations between them is difficult. At the same time, collected datasets may be noisy or so large that they become computationally challenging to process efficiently~\cite{liu2020large,pujara2017sparsity}.

One possible example to examine algorithms performance on incomplete data is to connect the study of spreading phenomena with graph reduction techniques. Most existing graph reduction benchmarks focus on single-relationship graphs and compare algorithms across diverse datasets~\cite{gong2024gc4nc,liu2025gcondenser}. In contrast, studies evaluating the influence maximisation problem often omit the impact of incomplete data due to challenges in data availability and rarely reflect the complex nature of real-world systems~\cite{zareie2023influence}.

For these reasons, we propose the \textit{Spreading-Oriented Reduction Benchmark} (\textit{SORB}), which evaluates diverse real-world networks across several variants of the influence maximisation problem, with particular attention to graph reduction as a preprocessing step. Our framework places graph reduction at the centre of the evaluation by standardising datasets (including both homogeneous and heterogeneous networks), diffusion settings, and representative tasks. This framework allows us to measure how different sparsification and coarsening strategies affect the predictions of influence maximisation (IM) models in both multilayer and single-layer networks. In this way, we shift the perspective from testing IM algorithms in isolation to assessing the suitability of reduction methods for supporting accurate and efficient analysis under realistic network conditions.
The main contributions of this work are the following findings: 
\begin{enumerate}
    \item  Positive impact of graph sparsification on seed set's $Gain@k$ for single-layer networks, in contrast to the slight $\mathrm{AUC}_{\mathrm{cutoff}}$ degradation observed for multilayer networks after flattening,
    \item The degradation in downstream performance (measured in terms of $\mathrm{AUC}_{\mathrm{cutoff}}$ and $Gain@k$) is primarily driven by the choice of reduction strategy rather than the nominal reduction rate,
    \item A learning-based IM model (\textbf{ts-net}) benefits measurably from sparsification preprocessing on homogeneous networks, and lightweight heuristics show small but consistent gains in the same setting.
\end{enumerate}

\section{Preliminaries}
This section introduces the formal setting of the problem and situates it within the relevant research directions, namely influence maximisation, graph reduction, and multilayer network modelling. To the best of our knowledge, there is no standardised framework for evaluating the impact of graph reduction on influence maximisation tasks, which motivates the proposed \textit{SORB}.

\subsection{Multilayer Networks}
Multilayer networks, as a partially overlapping class of heterogeneous graphs~\cite{Shi2022HeterogeneousGraphs}, provide a more realistic representation of complex systems by modelling multiple types of interactions between actors~\cite{jankowski2024timik}. Recently, research has shifted from homogeneous to more complex network models, opening new directions for analysis. In the context of spreading processes, examples include heterogeneous influence maximisation methods that leverage meta-path-based representations~\cite{Deng2020,mahe2022b}, as well as multilayer influence maximisation approaches that model diffusion across multiple interconnected layers~\cite{Venkatakrishna2022CIM}.

Multilayer networks extend the basic network model $G=(V, E)$, where $V$ is the set of nodes and $E$ is the set of edges. In this work, we adopt the multilayer network notation proposed by~\cite{kivela2014multilayer}, summarised in Def.~\ref{def:multilayer_net}.

\begin{definition}[Multilayer network]
    \label{def:multilayer_net}
    A multilayer network is defined as a quadruple 
    $M = (A,L,V,E)$, with:
    \begin{itemize}
        \item $A$ is a finite set of actors,
        \item $L$ is a finite set of layers,
        \item $V \subseteq A \times L$ is a set of nodes,
        \item $E \subseteq \bigcup_{l \in L} (V_l \times V_l)$ is a set of edges.
    \end{itemize}
\end{definition}

The main purpose of multilayer networks is to represent the same actor in the context of different types of relations that may occur in a system. This concept partially overlaps with the strict definition of heterogeneous graphs~\cite{Shi2022HeterogeneousGraphs}; however, multilayer networks simplify the representation of the actor set by omitting explicit distinctions between different classes of nodes. A multilayer network can therefore be viewed as a collection of graphs $G_l = (V_l, E_l)$ for $l \in L$ spanned on different subsets of actors. Each actor may be represented by at most one node per layer, i.e., $V \subseteq A \times L$. Edges are allowed only between nodes within the same layer (no interlayer edges). All edges are assumed to be undirected and unweighted, and neither nodes nor edges are associated with any features.

\subsection{Graph Reduction}
Graph reduction techniques, including sparsification and coarsening, have been widely studied with the aim of preserving selected properties of the original graph~\cite{hashemi2024comprehensive}. These methods establish a mapping between nodes in the original and reduced graphs. In contrast to graph condensation, their output is typically a reduced subgraph that either preserves structural information or aggregates similar nodes. Due to the broad range of applications of graph reduction---such as visualisation~\cite{zhao2018nearly}, privacy preservation~\cite{dong2022privacy}, and data augmentation~\cite{zhao2022graph}---their adaptation to more complex multirelational settings remains insufficiently explored. Graph reduction for multirelational graphs has been examined using approaches such as R-GCN-based methods~\cite{generale10scaling}; however, the interaction between graph reduction and spreading processes in such settings remains largely unexplored.

Graph reduction is defined as a mapping
\begin{equation}
    G' = \mathcal{R}(G), 
    \quad |V'| \leq |V|,\; |E'| \leq |E|.
\end{equation}
In this work, we adopt the taxonomy of graph reduction techniques introduced in~\cite{hashemi2024comprehensive}, summarised in Def.~\ref{def:reduction_taxonomy}.

\begin{definition}[Graph reduction]\label{def:reduction_taxonomy}
Graph reduction can be formulated as the process of finding a reduced graph $G'$ that minimises a loss function measuring the discrepancy between the original graph $G$ and its reduced counterpart:
\begin{equation}
    G' = \arg\min_{G'} \mathit{Loss}(G, G').
\end{equation}
\end{definition}

In the experiments, we consider two types of graph reduction described in Def.~\ref{def:graph_sparsification} (sparsification) and Def.~\ref{def:graph_coarsening} (coarsening).

\begin{definition}[Graph sparsification]\label{def:graph_sparsification}
Given a graph $G = (\mathbf{A}, X)$, where $\mathbf{A}$ is the adjacency matrix and $X$ is the node feature matrix, graph sparsification constructs a reduced graph $G' = (\mathbf{A}', X')$ by selecting a subset of the structural elements of the original graph. In practice, this corresponds to retaining a subset of edges (and optionally features) whilst preserving the original node set. Formally, the elements of $\mathbf{A}'$ or $X'$ constitute subsets of the corresponding elements in $\mathbf{A}$ or $X$. In our setting, the adjacency matrix retains its original dimensions.
\end{definition}

\begin{definition}[Graph coarsening]\label{def:graph_coarsening}
Given a graph $G = (\mathbf{A}, X)$, graph coarsening produces a reduced graph $G' = (\mathbf{A}', X')$ consisting of $N'$ super-nodes and $E'$ super-edges, where $N' < N$ denotes the number of nodes in the original graph. The procedure requires an assignment matrix $C \in \{0,1\}^{N \times N'}$ that associates each node in the original graph with a super-node in the coarsened graph.
\end{definition}

For influence maximisation tasks, prediction on the coarsened graph proceeds in two stages. First, the IM model generates predictions on the coarsened graph and selects a set of super-nodes. Second, each selected super-node is expanded to the original nodes it represents, which are internally ranked according to their degree in the original graph. This simple yet effective strategy has been shown to achieve strong performance in previous studies~\cite{czuba2025rank,czuba2025identifying}.

Due to the research gap in dedicated reduction techniques for diverse types of graphs~\cite{hashemi2024comprehensive}, it was necessary to apply an additional flattening transformation to multilayer networks, as described in Def.~\ref{def:flattening_network}.

\begin{definition}[Flattening transformation] 
\label{def:flattening_network} 
Let $M = (A, L, V, E)$ be a multilayer network as defined in Def.~\ref{def:multilayer_net}, and assume the underlying multilayer graph is undirected, so that$((a, l), (b, l)) \in E \iff ((b, l), (a, l)) \in E$. The flattening transformation $F : M \to G$ produces an undirected multigraph $G = (A, E_F)$ over the actor set $A$, with edge set 
\begin{equation} 
E_F = \bigl\{ (\{a, b\}, l) \mid a, b \in A,\; l \in L,\; ((a, l), (b, l)) \in E \bigr\}. 
\end{equation} 
Each element of $E_F$ pairs an unordered actor pair $\{a, b\}$with the layer $l$ from which the corresponding edge originates; parallel edges arising from different layers are therefore retained as distinct elements of $E_F$. Writing $E_l := \{\{a, b\} \mid ((a, l), (b, l)) \in E\}$ for the (undirected) edges of layer $l$, we have $|E_F| = \sum_{l \in L} |E_l|$. 
\end{definition}

The flattening transformation should be interpreted as a pragmatic preprocessing layer rather than a theoretically optimal multilayer reduction strategy. It enables the application of existing graph reduction algorithms, which are primarily designed for single-layer graphs, whilst preserving actor-level aggregation of interactions. This introduces potential edge duplication, which we explicitly analyse in Sec.~\ref{app:graph_reduction_impact}. We treat flattening as a controlled trade-off between data handling and computational feasibility of the model.

\subsection{Influence maximisation}
Introduced by~\cite{kempe2003maximizing}, influence maximisation was originally formulated as a combinatorial optimisation task. Its objective is to identify a set of agents that, when selected as initial nodes, maximise the spread of a process across the network. The majority of existing approaches rely on centrality-based heuristics, either directly or through their extensions, such as discounting strategies~\cite{chen2009efficient}. Recent works have highlighted the limited generalisation capabilities of such methods and proposed approaches based on latent node embeddings~\cite{kumar2022} or graph neural networks~\cite{ling2023deepim,czuba2025identifying}. Although considerable progress has been made and numerous seed selection methods have been proposed, the problem remains far from being fully resolved. Its complexity is further amplified by the continuous emergence of new network models and by the ongoing reformulation of the problem driven by its broad practical significance.

\subsection{Benchmarking}
Benchmarking graph reduction methods requires evaluating two key aspects: the reduction paradigm itself and its impact on downstream tasks. Recent works have primarily focused on graph condensation, evaluating its effectiveness on tasks such as node classification and graph classification~\cite{gong2024gc4nc,sun2024gc}. In the context of influence maximisation, benchmarking efforts have mainly addressed efficiency and scalability, aiming to enable the application of these methods to large-scale, realistic networks~\cite{arora2017debunking,arora2019influence}. Such efforts are well motivated, as realistic applications--ranging from epidemic containment on country-scale contact networks to reliability analysis of large infrastructure systems--typically involve graphs whose size renders direct analysis computationally infeasible, making graph reduction a practical necessity for deploying IM methods at scale.

\section{Experimental Setup and Results}
\label{app:experiments}
\begin{figure*}[ht!]
    \centering
    \includegraphics[width=\linewidth]{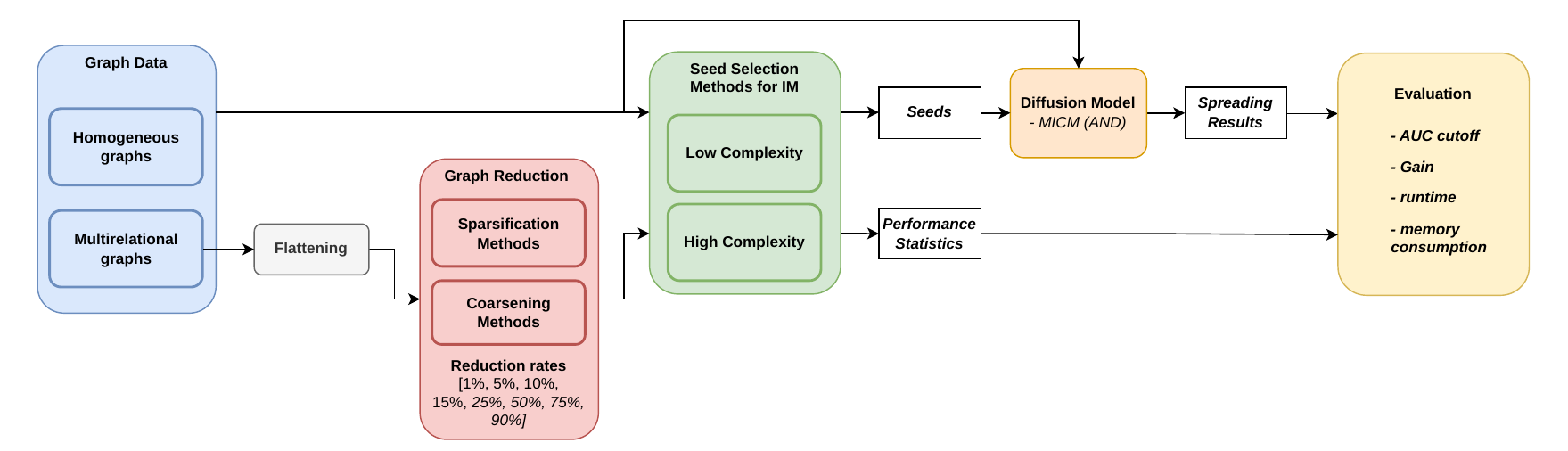}
    \caption{A schematic illustration of the \textit{SORB} processing pipeline. The main components are: the reduction module, which preprocesses the input network; the prediction module, which executes the seed selection models specified in the configuration on both original and reduced graphs; the diffusion module, which generates the spreading results for each predicted seed set by running MICM simulations on the original network; and the evaluation module, which consolidates the simulation outcomes into downstream task metrics and seed selection methods performance statistics, and prepares them for comparison.}
    \label{fig:sorb-pipeline}
\end{figure*}

The processing workflow of this study is organised into three stages and illustrated in Fig.~\ref{fig:sorb-pipeline}. In the first stage, the reduction module produces reduced versions of each network across the full grid of reduction rates and methods. In the second stage, each seed selection model is executed on every available version of the network (original and reduced), yielding a Cartesian product of networks, reduction methods, reduction factors $r$, and seed selection models for each downstream task. In the final stage, the diffusion module runs Multilayer Independent Cascade Model (\textbf{MICM}) simulations~\cite{czuba2025identifying} on the original network for every predicted seed set (regardless of whether it was produced on an original or a reduced graph), and the evaluation module consolidates the simulation outcomes into the downstream task metrics, computing the differences between scores obtained on the original and on the reduced networks.

To guide the design of the benchmark and the experimental evaluation, we formulated the following research questions: \textbf{Q1:} What is the impact of graph reduction on the performance of predicted seed sets of different sizes? \textbf{Q2:} How does the reduction rate $r$ affect downstream results? \textbf{Q3:} Does the application of graph reduction methods lead to consistent effects across diverse input networks? \textbf{Q4:} What is the impact of different graph reduction methods on the computational resources required by IM methods during prediction? All discussed experiments were conducted using the specification described in App.~\ref{app:training_details}.

\subsection{Datasets}
\label{app:dataset}
The datasets presented in Tab.~\ref{tab:networks} were selected to capture the diversity of real-world networks in terms of their topological characteristics, such as size, density, and number of relationships or layers. Our primary focus is on multilayer networks (MLNs), as they represent the richer and more realistic setting for spreading phenomena. Nevertheless, for comparison and to provide a reference point for scenarios without layered structure, the collection is also enriched with homogeneous graphs, namely \textit{WikiCS}~\cite{mernyei2020wiki} and \textit{Amazon-CS}~\cite{shchur2018pitfalls}. The multilayer side covers \textit{Freebase}~\cite{wang2021self}, \textit{IMDB}~\cite{wu2016explaining}, \textit{FinDKG}~\cite{lifindkg}, \textit{arxiv}~\cite{dedomenico2015arxiv}, and \textit{timik}~\cite{jankowski2024timik}, spanning a wide range of sizes, densities, and average degrees per actor.

For each actor in each network, we determined their spreading potential through simulations of the \textbf{MICM}~\cite{czuba2025identifying} under varying spreading parameters. We focus on the protocol function $AND$ and consider activation probabilities $\pi \in \{0.80, 0.85, 0.90, 0.95\}$. This choice is motivated by the need to impose demanding conditions for achieving high evaluation scores, in line with observations reported in~\cite{czuba2025rank,czuba2025identifying}. Due to the probabilistic nature of this spreading model, each simulation scenario was repeated $30$ times to obtain a statistically reliable sample. This step allows us to estimate the spreading potential of individual actors, as in~\cite{czuba2025identifying}, and to use these estimates for evaluating rankings and single-actor seed sets. Any additional \textbf{MICM} simulations required later by the evaluation module are performed exclusively on the original networks; in particular, when a seed set is predicted on a reduced graph, it is still evaluated by running the simulation on the corresponding original network.

\begin{table}[ht!]
    \centering
    \caption{Networks used in the \textit{SORB} with their key parameters: number of layers, actors, nodes, edges, and the average degree per actor.}
    \label{tab:networks}
    \addtolength{\tabcolsep}{-0.5em}
    \begin{tabular}{lrrrrr}
    Network & Layers & Actors & Nodes & Edges & AVG. Degree \\ \hline

    WikiCS~\cite{mernyei2020wiki} & 1 & 11 367 & 11 367 & 216123  & 36.85 \\
    Amazon-CS~\cite{shchur2018pitfalls} & 1 & 13 752 & 13 752 & 245 861 & 35.76 \\
    \hline
    Freebase~\cite{wang2021self} & 3 & 3 492 & 7 435 & 142 144 & 75.41 \\
    IMDB~\cite{wu2016explaining} & 2 & 3 550 & 7 100 & 43 658 & 20.60 \\
    FinDKG~\cite{lifindkg} & 15 & 10 670 & 45 806 & 93 991 & 17.56 \\
    
    arxiv~\cite{dedomenico2015arxiv} & 13 & 14 065 & 26 796 & 59 026 & 8.39 \\
    timik~\cite{jankowski2024timik} & 3 & 61 702 & 102 247 & 875 191 & 28.37 \\
    \end{tabular}
\end{table}

\subsection{Network flattening}
\label{app:n_flattening}
\begin{figure}[ht!]
    \centering
    \includegraphics[width=\linewidth]{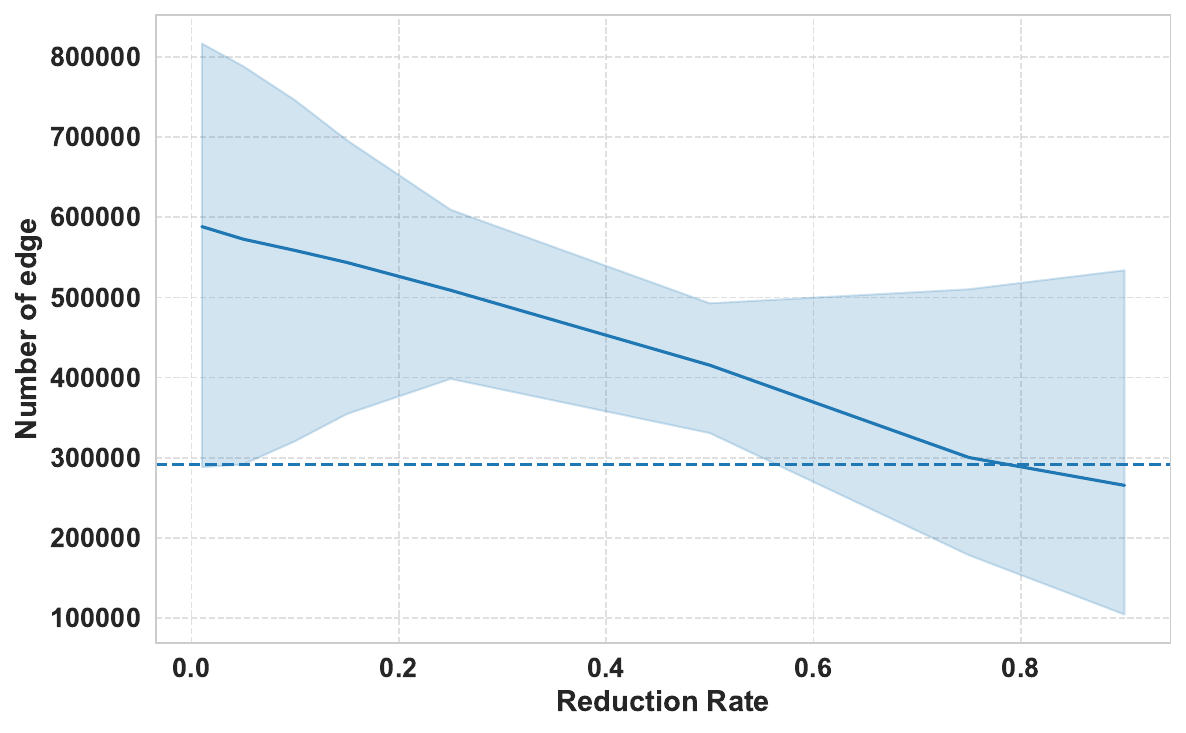}
    \caption{Number of edges in the sparsified versions of the \textit{timik} network. The dashed line shows the per-layer average number of edges in the original network.}
    \label{fig:per_layer_edges}
\end{figure}

Because the majority of considered reduction algorithms operate on homogeneous graphs, multilayer networks have been flattened into a single-layer representation before reduction can be applied, as described in Def.~\ref{def:flattening_network}. This transformation aggregates inter-layer relations between vertices into a single set of edges, enabling a uniform treatment of multilayer and homogeneous datasets within the pipeline. A direct consequence, however, is that many edges are duplicated across layers and subsequently collapsed, which substantially alters the structural profile of the network and carries implications for both downstream performance and resource consumption. An example of how the edge count changes after flattening and sparsification is shown in Fig.~\ref{fig:per_layer_edges} for the \textit{timik} network, which contains the largest number of edges in our collection; the dashed line denotes the per-layer average number of edges in the original graph. Notably, the flattened-and-sparsified edge count only approaches the per-layer average at the most aggressive setting ($r = 0.9$) and remains several times larger than a single original layer across the rest of the sweep, so even strong
single-layer sparsification cannot, on its own, undo the duplication introduced by flattening. This is a structural manifestation of the same overhead that we revisit later in the context of memory consumption for \textbf{ts-net} on heterogeneous graphs (Sec.~\ref{sec:im_methods}), and it reinforces our broader argument that the multilayer case calls for dedicated reduction algorithms rather than the flatten-then-reduce pipeline used here.

A natural alternative would be to apply each reduction method to every layer independently, which would limit the structural distortion introduced by flattening. We verified this intuition empirically on the \textit{timik} network which is the largest and most complex dataset in our collection as shown in Fig.~\ref{fig:timik_per_layer_vs_flattened} in App.~\ref{app:additional_results}. The comparison confirms that per-layer processing yields only modest improvements over the flattened setting which is most visibly for \textbf{TSpanner} in terms of median and for \textbf{Jaccard similarity} in terms of variance, while the overall ordering of reduction methods is preserved. We nonetheless deliberately depart from the per-layer strategy and evaluate reduction methods on the flattened, whole-network representation. The motivation is methodological. None of the reduction methods considered here was designed with multilayer structure in mind, and applying them per layer would mean evaluating each method on a structurally simpler input than the one it is meant to handle: a single layer captures neither the cross-layer co-occurrence of actors nor the cumulative connectivity that each actor exhibits across the network as a whole. Flattening into a multigraph over the actor set $A$, in the sense of Def.~\ref{def:flattening_network}, preserves both: every edge of $M$ contributes a corresponding edge to $G$, so a pair of actors connected in $k$ layers gives rise to $k$ parallel edges and each method is exposed to the full set of incidences of each actor. Running the reduction methods on the flattened graph, rather than layer by layer, is therefore, in our view, a more informative test of how they behave on multilayer inputs. We acknowledge that the empirical difference between the two strategies on \textit{timik} is small, which on its own does not force a choice between them; the flattened setting is preferable here precisely because it does not inflate performance artificially and therefore gives a more honest picture of what existing methods can offer on multirelational data. By forcing the reduction methods to confront the full complexity of the multilayer graph at once, our benchmark makes explicit that no current method is particularly well-suited to this setting, thereby directly motivating the need for dedicated multilayer reduction algorithms. The difficulties observed in the following sections should be read in this light.

\begin{table}[ht]
  \centering
  \caption{Mean average degree per network and reduction method, averaged across all tested reduction rates $r$.}
  \label{tab:avg_degree}
  \resizebox{1\linewidth}{!}{%
  \addtolength{\tabcolsep}{-0.4em}
  \begin{tabular}{l|c|c|c|c|c|c}
    \toprule
    \textbf{Network} & \textbf{Original} & \textbf{ForestFire} & \textbf{Jaccard similarity} & \textbf{LocalDegree} & \textbf{Random} & \textbf{TSpanner} \\
    \midrule
    WikiCS & 37.94 & 16.01 & 25.75 & 25.97 & 25.11 & 16.57 \\
    Amazon-CS & 36.50 & 15.11 & 21.46 & 24.92 & 24.15 & 15.93 \\
    Freebase & 73.99 & 26.35 & 47.82 & 51.35 & 48.96 & 32.14 \\
    IMDB & 21.07 & 9.27 & 14.16 & 14.55 & 13.94 & 9.76 \\
    FinDKG & 12.92 & 6.08 & 7.85 & 8.76 & 8.55 & 5.87 \\
    arxiv & 5.00 & 2.46 & 3.23 & 3.83 & 3.31 & 2.93 \\
    timik & 26.67 & 11.77 & 16.62 & 18.48 & 17.64 & 11.56 \\
    \bottomrule
  \end{tabular}
  }
\end{table}

The structural footprint of flattening combined with
reduction is summarised in Tab.~\ref{tab:avg_degree}, which reports the mean average degree per network and reduction method,
averaged across all tested reduction rates~$r$. Notably, \textbf{ForestFire} reduces the average degree most aggressively
across the majority of networks, with \textbf{TSpanner} following
closely and exceeding it only on \textit{Freebase}. The two methods
therefore form a clearly separated pair of strong reducers, whereas
\textbf{Jaccard similarity}, \textbf{LocalDegree}, and \textbf{Random}
retain roughly two-thirds of the original degree, consistent with
their smaller effect sizes observed later in the evaluation. To illustrate the dependence on~$r$ without
aggregation, Tab.~\ref{tab:avg_degree_timik} presents a per-rate breakdown of the average degree for the \textit{timik} network. The breakdown reveals that the picture is highly non-uniform across methods: \textbf{TSpanner} exhibits a counter-intuitive, monotonically increasing trajectory with~$r$, collapsing the average degree to $\approx 2.01$ at $r = 0.01$ but leaving it essentially intact ($25.91$, against an original value of $26.67$) at $r = 0.9$. This behaviour reflects how our implementation couples~$r$ to the spanner construction: $r$ acts as a testing budget for a randomised, $t = 4$ stretch criterion applied to edges removed from a working copy of the graph, and only stretch-critical edges among those tested are retained (see App.~\ref{app:tspanner_adapt} for details). Larger~$r$ therefore exposes more edges to the test and accumulates more critical ones in the output, which is the opposite of the convention used by every other reduction method in our pool. \textbf{ForestFire}, by contrast,
behaves almost as a step function, holding around~$14$ up to $r = 0.5$
and then dropping sharply to about~$5$; whereas \textbf{Random},
\textbf{LocalDegree}, and \textbf{Jaccard similarity} decrease
gradually with~$r$, in the direction one would intuitively expect. A
non-trivial crossover also emerges at extreme reduction rates: at
$r = 0.9$, \textbf{Random} edge removal leaves a lower average degree
($2.68$) than the targeted \textbf{LocalDegree} strategy~($3.25$),
suggesting that when only a small fraction of edges survives,
untargeted sampling can become structurally more destructive than
hub-preserving alternatives. This heterogeneity, and in particular
TSpanner's inverted dependence on~$r$, foreshadows our later
observation that the choice of reduction strategy dominates
over the nominal reduction rate.

\begin{table}[ht]
  \centering
  \caption{Average degree for the \textit{timik} network per reduction method and rate $r$. The original network has an average degree of 26.67.}
  \label{tab:avg_degree_timik}
  \resizebox{1\linewidth}{!}{%
  \addtolength{\tabcolsep}{-0.4em}
  \begin{tabular}{l|c|c|c|c|c}
    \toprule
    \textbf{$r$} & \textbf{ForestFire} & \textbf{Jaccard similarity} & \textbf{LocalDegree} & \textbf{Random} & \textbf{TSpanner} \\
    \midrule
    0.01 & 14.03 & 26.30 & 26.67 & 26.40 & 2.01 \\
    0.05 & 13.98 & 24.79 & 26.08 & 25.34 & 2.66 \\
    0.10 & 13.95 & 22.93 & 25.00 & 24.01 & 4.68 \\
    0.15 & 14.07 & 21.06 & 23.81 & 22.67 & 6.54 \\
    0.25 & 13.99 & 17.34 & 21.20 & 20.02 & 9.97 \\
    0.50 & 13.97 & 8.29 & 14.38 & 13.34 & 17.39 \\
    0.75 & 5.10 & 6.14 & 7.48 & 6.66 & 23.31 \\
    0.90 & 5.07 & 6.14 & 3.25 & 2.68 & 25.91 \\
    \bottomrule
  \end{tabular}
  }
\end{table}

Addressing \textbf{Q3}, we further examined which structural properties of the flattened networks correlate with downstream performance changes. Tab.~\ref{tab:correlation} presents Spearman correlations between structural factors and the observed degradation of $\text{AUC}_{\text{cutoff}}$ (described in detail in Def.~\ref{def:actor_ranking}). Among the examined structural features, only density demonstrates a statistically significant correlation with degradation ($\rho = -0.160$, $p = 0.007$), and the sign is negative: denser flattened networks tend to suffer less from reduction. A plausible reading is that, when many alternative diffusion pathways are available, removing a fraction of them leaves the influence ranking comparatively intact, whereas sparser networks rely on a smaller number of critical edges whose loss is harder to absorb. By contrast, neither average degree ($\rho = 0.037$, $p = 0.537$) nor degree variance ($\rho = -0.090$, $p = 0.134$) reaches significance, indicating that the local concentration of connectivity is not on its own a reliable predictor of ranking stability. Overall, flattening should therefore be treated as a non-trivial structural operation whose trade-off cost must be accounted for when interpreting reduction results.

\begin{table}[t]
    \centering
    \caption{Spearman correlation between structural factors and graph reduction.}
    \label{tab:correlation}
    \begin{tabular}{lcc}
    \toprule
    Variable & $\rho$ & $p$-value \\
    \midrule
    Reduction rate $r$ & -0.058 & 0.331 \\
    Density & -0.160 & 0.007 \\
    Average degree & 0.037 & 0.537 \\
    Degree std & -0.090 & 0.134 \\
    \bottomrule
    \end{tabular}
\end{table}

\subsection{Graph reduction methods}
\label{app:graph_reduction_impact}

For the graph reduction module, we build upon the interface introduced in~\cite{gong2025gcnc} and incorporate selected algorithmic implementations from~\cite{angriman2023algorithms}. We adopt well-established and strong baselines that are widely used in the literature and serve as competitive reference points for graph reduction~\cite{gong2024gc4nc}. These methods were selected to cover a wide range of reduction paradigms, including random, structure-preserving, and similarity-based strategies. As a baseline, we employ sparsification based on \textbf{Random} edge sampling, where edges preserved in the sparsified graph are sampled uniformly at random. We further consider several established sparsification techniques: the \textbf{LocalDegree} method prioritises edges connected to high-degree nodes, following the intuition that hubs play a central role in network structure; the \textbf{Jaccard similarity} approach~\cite{satuluri2011local} selects edges based on the similarity between the neighbourhoods of their endpoints (note that the name refers to the edge-scoring function and is unrelated to the Jaccard index later used within the $\mathrm{AUC}_{\mathrm{cutoff}}$ metric); the \textbf{TSpanner} method~\cite{liestman1993additive} retains a subset of edges that approximately preserves shortest-path distances in the graph; in our experiments we use an adaptation of this construction that exposes a single reduction rate~$r$, described in App.~\ref{app:tspanner_adapt}; and \textbf{ForestFire}~\cite{leskovec2007graph}, a stochastic sparsification technique based on a recursive edge-burning process. For graph coarsening, we consider two representative methods: \textbf{VariationNeighborhoods}~\cite{loukas2019graph,huang2021scaling}, which aggregates nodes with similar neighbourhood structures, and \textbf{AffinityGs}~\cite{gong2025gcnc}, which performs greedy matching by iteratively selecting edges with the highest affinity and removing adjacent edges from further consideration. Each method applies a reduction controlled by a factor $r$, where $r \in \{0.01, 0.05, 0.1, 0.15, 0.25, 0.5, 0.75, 0.9\}$.

\begin{figure}[ht!]
    \centering
    \includegraphics[width=\linewidth]{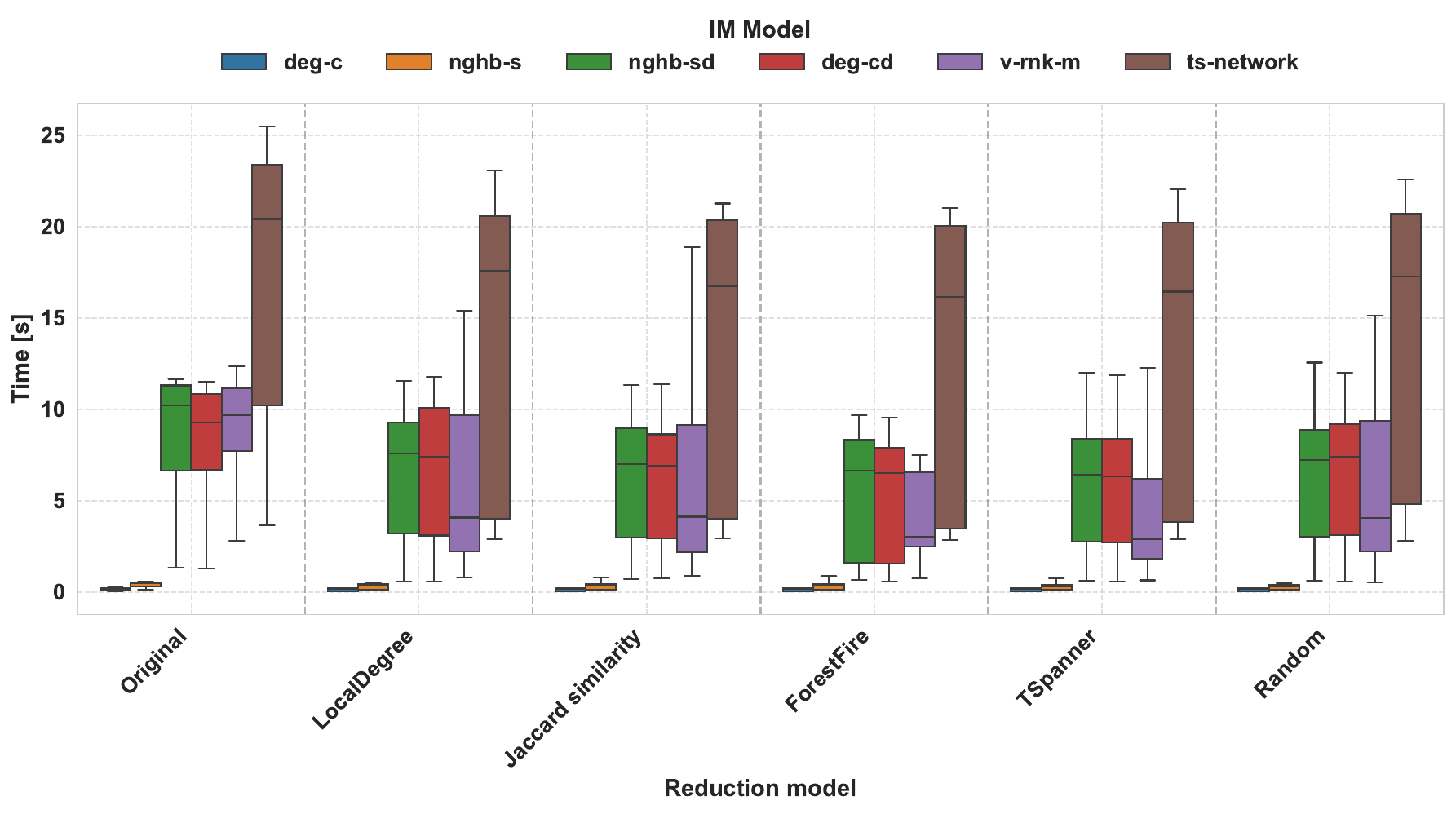}
    \caption{Influence maximisation model execution times with the applied reduction model.}
    \label{fig:time_heuristics}
\end{figure}

\begin{figure*}[ht!]
    \centering
    \includegraphics[width=\linewidth]{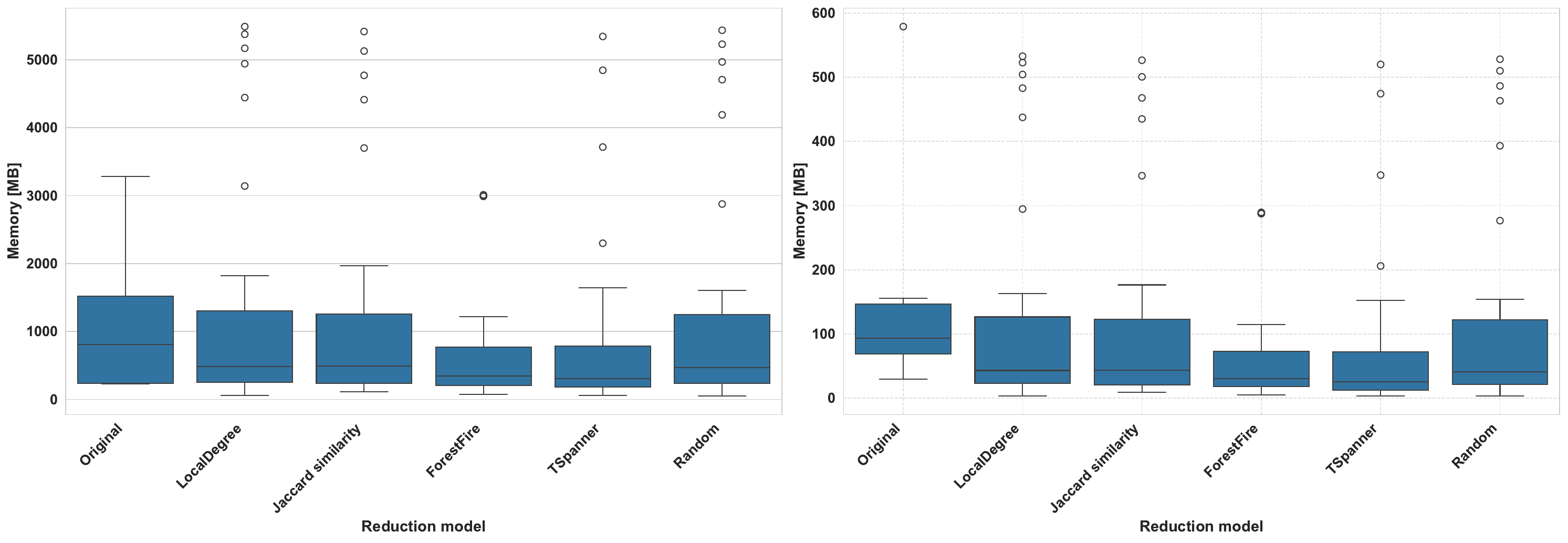}
    \caption{Memory usage during IM model execution. Left panel: peak GPU memory consumption (MB) for \textbf{ts-net}. Right panel: peak RAM usage (MB) for \textbf{v-rnk-m}; its GPU memory consumption was approximately an order of magnitude lower and is omitted. Note that the two panels use independent y-axis scales.}
    \label{fig:memory-comparison}
\end{figure*}

A practical consequence of these design differences is that coarsening and sparsification cannot be evaluated on equal footing in terms of computational cost. In our experiments, coarsening was consistently slower in wall-clock time by a large margin on every network for which we attempted it, to the point that completing the full sparsification-style sweep (seven reduction rates $\times$ six IM models $\times$ all seven networks) was infeasible within the computational budget allocated to this study. Exploratory runs on smaller networks did not surface performance or memory improvements that would justify the investment required to close this gap. For this reason, the time and memory analysis below focuses exclusively on sparsification, whereas coarsening is revisited later only in the context of seed-set $Gain$ on the subset of networks where it was feasible to run--namely \textit{AmazonCS}, \textit{WikiCS}, \textit{Freebase}, and \textit{arxiv}.

\begin{table}[t]
    \centering
    \caption{Statistical analysis of performance impact on $\mathrm{AUC}_{\mathrm{cutoff}}$ for applied graph reduction methods.}
    \label{tab:significance}
    \resizebox{1\linewidth}{!}{%
    \addtolength{\tabcolsep}{-0.4em}
    \begin{tabular}{lccc}
    \toprule
    Method & Wilcoxon $p$-value & Cohen's $d$ & Effect Size \\
    \midrule
    ForestFire & $6.7\times 10^{-6}$  & $-0.66$ & Moderate--Large \\
    Jaccard similarity & $2.3\times 10^{-4}$  & $-0.47$ & Small--Moderate \\
    TSpanner & $8.3\times 10^{-4}$  & $-0.42$ & Small--Moderate \\
    Random & $1.31\times 10^{-2}$ & $-0.36$ & Small \\
    LocalDegree & $1.13\times 10^{-2}$ & $-0.32$ & Small \\
    \bottomrule
    \end{tabular}
    }
\end{table}

Turning now to the computational side of sparsification, and directly addressing \textbf{Q4}, inference time on the reduced graphs, illustrated in Fig.~\ref{fig:time_heuristics}, decreases on average across all IM models after applying any sparsification algorithm, with the most pronounced improvements observed for the heavier methods. Memory consumption follows the same trend for the two resource-critical IM models, \textbf{ts-net} and \textbf{v-rnk-m}, as presented in Fig.~\ref{fig:memory-comparison}: the intuitive expectation is confirmed, as smaller networks are easier to store and process, and we observe noticeably lower memory requirements for all reduction models, both in terms of mean values and the upper ranges of the box plots. The remaining IM models exhibit negligible memory footprints and are therefore omitted from this comparison.

\subsubsection{Significance of reduction effects}
To quantify the statistical significance of reduction effects, we applied the paired Wilcoxon signed-rank test comparing performance on reduced and original graphs. Tab.~\ref{tab:significance} shows that all reduction strategies result in statistically significant degradation of $\text{AUC}_{\text{cutoff}}$ ($p < 0.05$). \textbf{ForestFire} exhibits the strongest effect ($p = 6.7\times 10^{-6}$, $d = -0.66$), indicating a moderate-to-large practical impact. \textbf{Jaccard similarity} and \textbf{TSpanner} produce small-to-moderate degradation ($d = -0.47$ and $d = -0.42$, respectively), while \textbf{Random} and \textbf{LocalDegree} demonstrate smaller but still significant effects ($d = -0.36$ and $d = -0.32$). Importantly, even purely random edge removal induces statistically significant ranking shifts ($p = 1.31\times 10^{-2}$, $d = -0.36$), suggesting that influence ranking is inherently sensitive to structural perturbations.

Turning to \textbf{Q2}, we did not observe a significant monotonic relationship between reduction rate~$r$ and $\text{AUC}_{\text{cutoff}}$ loss ($\rho = -0.058$, $p = 0.331$). This formalises the threshold effect already anticipated in the previous subsection and indicates that degradation is primarily driven by the structural strategy of reduction rather than by the nominal percentage of removed edges. The per-rate decomposition in Tab.~\ref{tab:avg_degree_timik} provides a concrete illustration of this mechanism: \textbf{ForestFire} behaves almost as a step function, holding its average degree nearly constant up to $r = 0.5$ before dropping sharply, so that nominally different reduction rates can yield nearly identical structural outcomes; and \textbf{TSpanner}, conversely, moves in the opposite direction to the nominal rate, with low~$r$ corresponding to the most aggressive edge removal. In both cases the relationship between~$r$ and the realised structural impact is non-monotonic, which is exactly the pattern one would expect in the absence of a monotonic metric-level dependence on~$r$. This complements the structural-sensitivity observations reported earlier in the context of flattening.

\subsection{Influence maximisation methods}
\label{sec:im_methods}
During the study, we evaluated several models for predictions in influence maximisation tasks. These models represent different methodological categories and can be broadly classified into ranking-based approaches of varying complexity. The first group consists of commonly used centrality-based heuristics with low complexity, including Degree Centrality (\textbf{deg-c}), Degree Discount (\textbf{deg-cd})~\cite{chen2009efficient}, Neighbourhood Size (\textbf{nghb-s})~\cite{magnani2011ml}, and Neighbourhood Size Discount (\textbf{nghb-sd})~\cite{czuba2025rank}. These methods were selected due to their low computational cost and their ability to produce actor rankings compatible with the multilayer setting. Note that for monoplex networks \textbf{nghb-s} degenerates to \textbf{deg-c}, and analogously \textbf{nghb-sd} reduces to \textbf{deg-cd}; this is relevant for our evaluation, which spans both monoplex and multilayer networks. Another category of high complexity algorithms is represented by the multilayer extension of the VoteRank algorithm (\textbf{v-rnk-m})~\cite{czuba2025rank}. For learning-based approaches, we included the Top Spreader Network (\textbf{ts-net})~\cite{czuba2025identifying}, as an example of the inductive model. To analyse the impact of graph reduction on \textbf{ts-net}, we modified its implementation by disabling batching and additional preprocessing steps that remove isolated nodes. Particular attention is given to \textbf{ts-net} and \textbf{v-rnk-m}, as these methods are computationally more demanding by design, and therefore graph reduction may provide the greatest benefits in terms of reduced computational time and resource consumption.

In practice, the resource savings reported in the previous subsection translate differently across the IM families. The lightweight heuristics \textbf{deg-c} and \textbf{nghb-s} exhibit negligible execution times across all considered scenarios, so the savings enabled by reduction are most visible for the heavier methods: \textbf{v-rnk-m} combined with \textbf{TSpanner} achieves the greatest improvement in inference time, and memory consumption for both \textbf{v-rnk-m} and \textbf{ts-net} decreases with reduction, as shown in Fig.~\ref{fig:memory-comparison}. For the \textbf{ts-net} model, however, several outliers can be observed in which, even after applying reduction, more memory was required than on the original network. Such situations occur for certain sparsification settings with $r \leq 0.5$ across all heterogeneous graphs. They are a direct consequence of the flattening step described in Sec.~\ref{app:n_flattening}: because flattening preserves rather than collapses parallel edges originating from different layers (Def.~\ref{def:flattening_network}), the resulting single-layer representation contains strictly more edges than any individual layer, and for actor pairs connected across many layers this multiplies the per-edge cost carried by the \textbf{ts-net} pipeline. For low values of $r$, the subsequent sparsification is not aggressive enough to offset this overhead, which is why the anomaly is confined to $r \leq 0.5$ on heterogeneous graphs.

\subsection{Evaluation metrics and downstream results}
\begin{table*}[ht!]
    \centering
    \caption{
        Average benefit/drop ($\mathrm{mean} \pm \mathrm{std}$) of the evaluated influence maximisation methods across homogeneous and heterogeneous networks. Values represent the relative change in $\mathrm{AUC}_{\mathrm{cutoff}}$ compared with the originally predicted ranking for each rate of network sparsification. The largest and smallest effects are highlighted in colour: \biggest{biggest}, \lowest{lowest}.
        }
    \resizebox{1\linewidth}{!}{%
    \addtolength{\tabcolsep}{-0.4em}

    \begin{tabular}{l|llllllllllll}
        \toprule
        \multirow{2}{*}{$r$} & \multicolumn{2}{c}{deg-c} & \multicolumn{2}{c}{deg-cd} & \multicolumn{2}{c}{nghb-s} & \multicolumn{2}{c}{nghb-sd} & \multicolumn{2}{c}{ts-net} & \multicolumn{2}{c}{v-rnk-m} \\
        \cmidrule(lr){2-3} \cmidrule(lr){4-5} \cmidrule(lr){6-7} \cmidrule(lr){8-9} \cmidrule(lr){10-11} \cmidrule(lr){12-13}
            & Heterogeneous & Homogeneous & Heterogeneous & Homogeneous & Heterogeneous & Homogeneous & Heterogeneous & Homogeneous & Heterogeneous & Homogeneous & Heterogeneous & Homogeneous \\
            \midrule
            0.01 & 0.00 ± 0.05 & 0.01 ± 0.01 & -0.02 ± 0.04 & 0.00 ± 0.00 & -0.02 ± 0.05 & 0.01 ± 0.01 & -0.02 ± 0.03 & 0.00 ± 0.00 & -0.05 ± 0.08 & 0.03 ± 0.04 & -0.01 ± 0.01 & 0.00 ± 0.01 \\
            0.05 & 0.00 ± 0.05 & 0.00 ± 0.01 & -0.02 ± 0.04 & 0.00 ± 0.00 & -0.02 ± 0.05 & 0.00 ± 0.01 & -0.02 ± 0.03 & 0.00 ± 0.00 & -0.05 ± 0.07 & 0.02 ± 0.03 & -0.01 ± 0.01 & 0.00 ± 0.01 \\
            0.10 & 0.00 ± 0.05 & 0.00 ± 0.01 & -0.01 ± 0.04 & 0.00 ± 0.01 & -0.02 ± 0.05 & 0.00 ± 0.01 & -0.01 ± 0.03 & 0.00 ± 0.01 & -0.06 ± 0.07 & 0.02 ± 0.03 & -0.01 ± 0.01 & 0.00 ± 0.00 \\
            0.15 & 0.01 ± 0.05 & 0.00 ± 0.02 & -0.01 ± 0.03 & 0.00 ± 0.01 & -0.01 ± 0.04 & 0.00 ± 0.02 & -0.01 ± 0.03 & 0.00 ± 0.01 & -0.05 ± 0.07 & 0.04 ± 0.04 & -0.01 ± 0.01 & 0.00 ± 0.01 \\
            0.25 & 0.01 ± 0.04 & 0.00 ± 0.02 & -0.01 ± 0.03 & -0.01 ± 0.01 & -0.01 ± 0.04 & 0.00 ± 0.02 & -0.01 ± 0.03 & -0.01 ± 0.01 & -0.05 ± 0.06 & 0.04 ± 0.04 & -0.01 ± 0.02 & -0.01 ± 0.01 \\
            0.50 & -0.01 ± 0.04 & 0.00 ± 0.02 & -0.02 ± 0.03 & 0.00 ± 0.01 & -0.02 ± 0.05 & 0.00 ± 0.02 & -0.02 ± 0.03 & 0.00 ± 0.01 & -0.05 ± 0.06 & 0.04 ± 0.04 & -0.02 ± 0.03 & -0.01 ± 0.01 \\
            0.75 & -0.03 ± 0.07 & 0.01 ± 0.02 & -0.04 ± 0.06 & 0.00 ± 0.01 & -0.04 ± 0.08 & 0.01 ± 0.02 & -0.04 ± 0.06 & 0.00 ± 0.01 & -0.05 ± 0.07 & 0.06 ± 0.05 & -0.03 ± 0.04 & -0.01 ± 0.01 \\
            0.90 & -0.04 ± 0.07 & 0.01 ± 0.01 & -0.05 ± 0.06 & 0.00 ± 0.01 & -0.05 ± 0.08 & 0.01 ± 0.01 & -0.05 ± 0.07 & 0.00 ± 0.01 & \lowest{-0.06} ± \lowest{0.06} & \biggest{0.07} ± \biggest{0.05} & -0.03 ± 0.05 & -0.01 ± 0.01 \\
        \bottomrule
    \end{tabular}
    }
    \label{tab:gain_loss_auc}
\end{table*}

The evaluation module computes two task-specific metrics and contrasts their values obtained on the reduced networks with those obtained on the original graphs. The two downstream tasks considered are seed set selection and ranking prediction, and together they cover the vast majority of the evidence reported in this study. It is worth noting that the two rankings reported in Tab.~\ref{tab:significance} and Tab.~\ref{tab:gain_loss_auc} operate on different axes. The former identifies the IM model most sensitive to reduction (\textbf{nghb-s}), while the latter identifies the reduction strategy with the largest structural impact (\textbf{ForestFire}) and they should be read as complementary rather than overlapping views of the same phenomenon. A practical consequence is that the safest combinations for downstream deployment on multilayer data are those that are robust on both axes simultaneously. Most notably \textbf{v-rnk-m} paired with \textbf{LocalDegree} or \textbf{Random}, where neither the IM model nor the reduction strategy contributes a large individual effect. Conversely, \textbf{nghb-s} combined with \textbf{ForestFire} represents the least favourable pairing in our evaluation.

\subsubsection{Seed set selection}
The influence maximisation problem, formalised in the seminal work of~\cite{kempe2003maximizing}, asks for a set of initial seed nodes whose activation maximises the spread of influence throughout the network. As the field has developed, researchers have focused on various aspects of seed selection, ranging from optimising seed selection strategies~\cite{singh2022influence}, through activation mechanisms~\cite{brodka2021sequential}, to budget constraints on the number of selected seeds~\cite{czuba2025rank}. In our study, we place particular emphasis on the latter aspect due to its natural synergy with the preprocessing step considered in our framework. This allows us to analyse how different seed budgets, including the most restrictive case of selecting only a single `super-spreader'~\cite{kitsak2010identification}, behave under various forms of network incompleteness. To this end, we compare the $Gain@k$ values obtained on the original network $G$ with those computed on the reduced network $G'$. The underlying intuition is that, if graph reduction preserves the structural properties relevant to diffusion, then a seed set selected on $G'$ should perform comparably to one selected on $G$ when both are evaluated on the original network $G$ under diffusion model. In order to evaluate the predicted seed sets produced by IM models, we adopt the $Gain@k$ ($g@k$) metric introduced in~\cite{czuba2025rank}, described in Def.~\ref{def:ss_gain}.

\begin{definition}[Seed set Gain]\label{def:ss_gain}
For a seed size $k$, set to either a single actor ($k=1$) or to $1\%$, $3\%$, or $5\%$ of the total number of actors $|A|$ in the network, the $Gain$ is computed as:
\begin{equation}
\mathrm{Gain} = 100 \cdot \frac{e - k}{|A| - k}
\end{equation}
where $e$ denotes the number of actors exposed by a utilised seed set of size $k$ (the computation procedure is described in Sec.~\ref{app:dataset}).
\end{definition}

\begin{figure*}[ht!]
    \centering
    \begin{minipage}{0.45\textwidth}
        \centering
        \includegraphics[width=\linewidth]{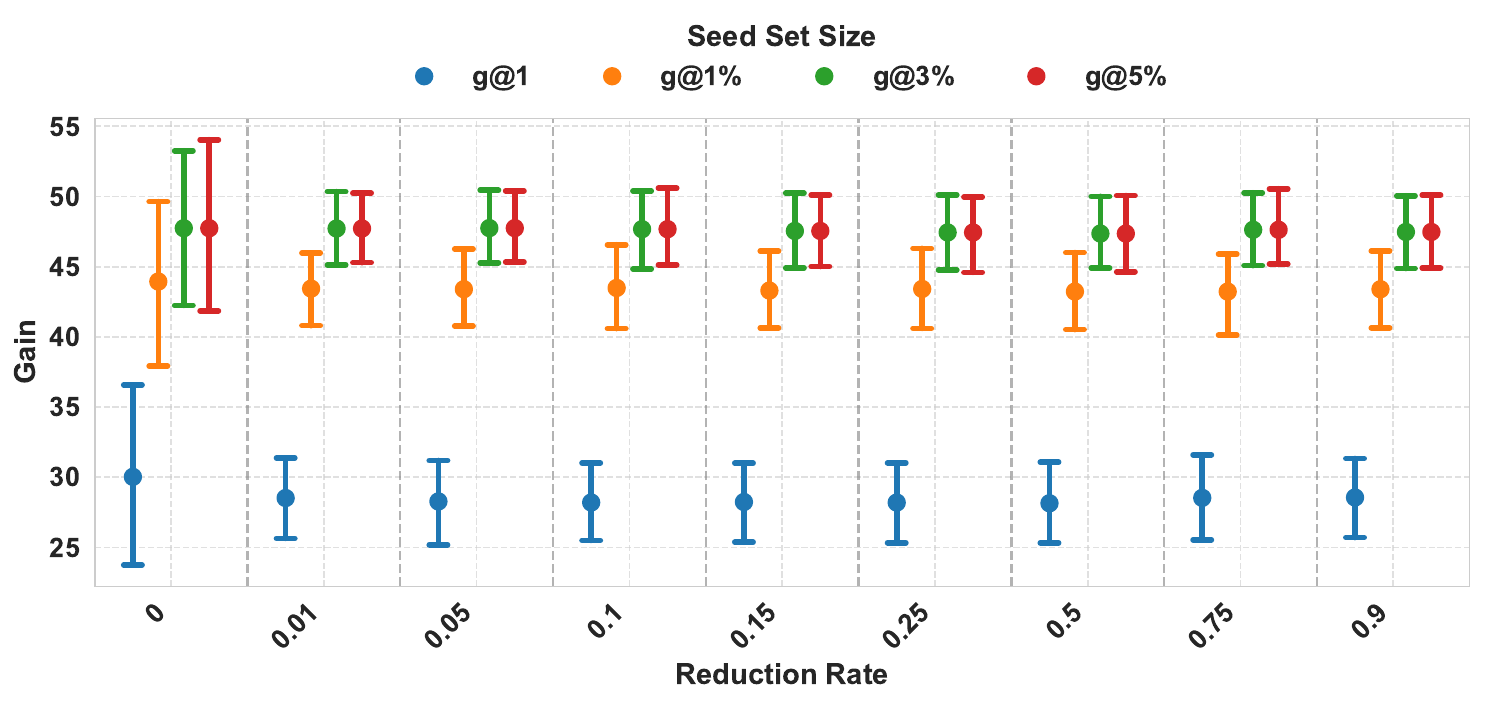}
    \end{minipage}
    \hfill
    \begin{minipage}{0.45\textwidth}
        \centering
        \includegraphics[width=\linewidth]{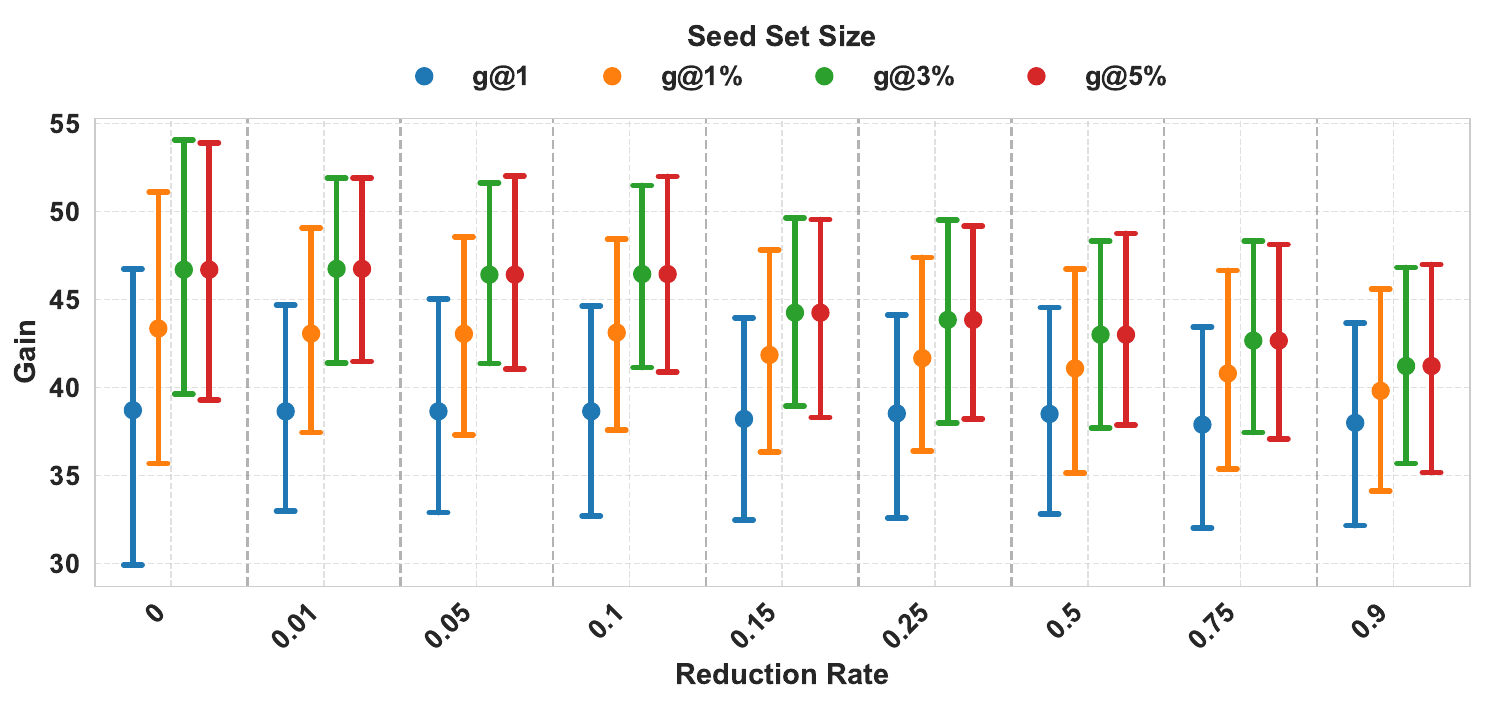}
    \end{minipage}
    \caption{Gain achieved for different seed sizes on examined networks using various reduction rates. A reduction rate of $0$ denotes the baseline, i.e. results obtained on the original network. On the left, applied sparsification models; on the right, coarsening models.}
    \label{fig:seed_set_gains_total}
\end{figure*}

\begin{figure}[ht!]
    \centering
    \includegraphics[width=\linewidth]{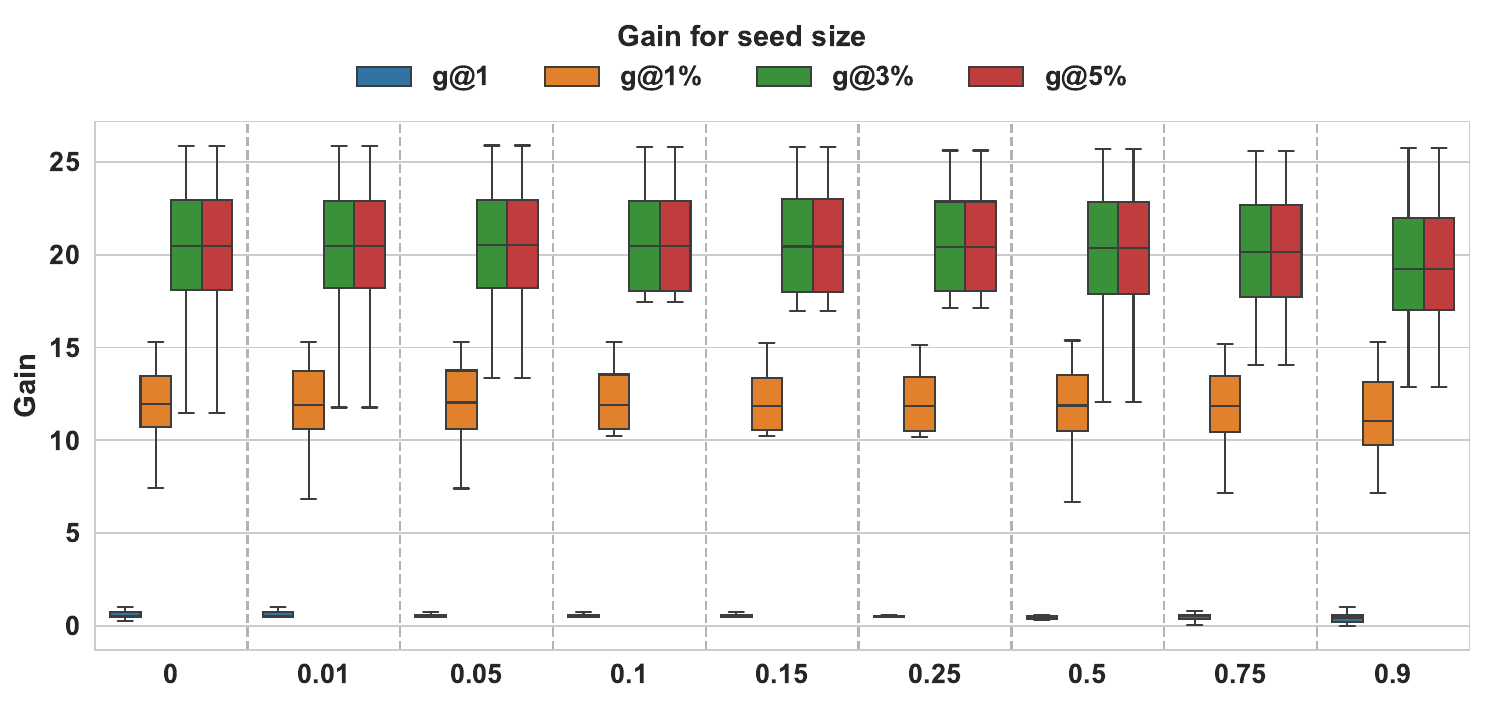}
    \caption{Gain score for \textit{FinDKG} network with the applied sparsification rate r.}
    \label{fig:findkg_gain_seed_set_rr}
\end{figure}

Directly addressing \textbf{Q1} and \textbf{Q3}, the difference in performance changes between the sparsification and coarsening methods is shown in Fig.~\ref{fig:seed_set_gains_total}. It is important to emphasise the difference in the data underlying the two panels: owing to the long processing times of the coarsening models and the lack of observable performance gains, we decided not to conduct experiments with coarsening on the full dataset, and instead limited the analysis to \textit{AmazonCS}, \textit{WikiCS}, \textit{Freebase}, and \textit{arxiv}. Although this does not constitute the complete dataset, the diversity of the selected networks allows us to observe a slight decline in the achieved $Gain$ values after coarsening for seed sets of all examined sizes. The decrease is marginal; for $Gain@1$, the average performance remains nearly unchanged, accompanied by lower variance. Moreover, variance consistently decreases across all seed sizes, indicating improved stability of the results, and the lower bounds of the achieved scores also remain above the baseline. The situation differs somewhat in the context of sparsification methods: the bounds of the achieved $Gain$ are considerably closer to each other across all cases, and the mean performance remains at the same level as the baseline. Importantly, as $r$ increases, we do not observe the progressive degradation in performance evident when preprocessing with coarsening models. A more detailed inspection of the results confirms this effect on an individual network level, for example \textit{FinDKG}, shown in Fig.~\ref{fig:findkg_gain_seed_set_rr}; a complete per-network breakdown of $Gain@k$ as a function of the reduction rate $r$, with seed size $k$ encoded by colour, is provided in App.~\ref{app:additional_results} and allows for a detailed comparison of the reduction impact across the different seed-size budgets. We may therefore conclude that, at the level of aggregate behaviour visible in Fig.~\ref{fig:seed_set_gains_total} and Fig.~\ref{fig:findkg_gain_seed_set_rr}, applying sparsification to the input networks at various levels does not appreciably deteriorate the selection of initial seed nodes, with means remaining at baseline and variance bounds tightening rather than widening.

\subsubsection{Ranking prediction}
With the aim of observing and analysing the cumulative spreading potential of actors, they can be ordered in a ranking according to their estimated influence, denoted as $R$, similarly to the approach proposed in~\cite{czuba2025identifying}. In this work, we refine this idea and provide a more formal presentation, as described in Def.~\ref{def:actor_ranking}. Importantly, the resulting rankings remain directly comparable, since both coarsening and sparsification in our setup are fully invertible transformations and preserve the underlying set of actors.

\begin{definition}[Area under curve at spreading cutoff]\label{def:actor_ranking}
Given a ground-truth ranking $\mathbf{R}$ and a predicted ranking $\hat{\mathbf{R}}$, 
the $\mathrm{AUC}_{\mathrm{cutoff}}$ metric measures the agreement between the rankings 
within the most influential portion of the network defined by the cutoff $c$. 
It is computed as:
\begin{equation}
\mathrm{AUC}_{\mathrm{cutoff}} = \frac{1}{n_c} \sum_{i=1}^{n_c - 1} \frac{J(i) + J(i+1)}{2}, 
\qquad n_c = \lfloor c|A| \rfloor
\end{equation}
where:
\begin{itemize}
    \item $J(k)$ -- Jaccard similarity between the top-$k$ prefixes of $\mathbf{R}$ and $\hat{\mathbf{R}}$,
    \item $A$ -- finite set of actors in the network,
    \item $c \in (0, 1]$ -- seed-set cutoff expressed as a fraction of $|A|$,
    \item $n_c$ -- number of actors within the cutoff; the prefactor $1/n_c$ normalises 
    the metric to $[0, 1]$.
\end{itemize}
\end{definition}

The cutoff is defined, following~\cite{czuba2025identifying}, as the 80th percentile of each network, restricting the analysis to the most influential actors. We evaluate ranking similarity using the Jaccard index computed over prefix sets of the two permutations, progressively from the top-$k$ elements to the full ranking. Evaluating rankings in this manner allows us to analyse not only whether the most influential actors are correctly identified, but also how well the relative ordering of high-impact nodes is preserved under different reduction strategies.

The ranking prediction scenario is particularly relevant for \textbf{Q2} and \textbf{Q3}. The averaged results of the predictions obtained using the influence maximisation models are presented in Tab.~\ref{tab:gain_loss_auc}, from which several observations can be highlighted. First, sparsification exhibits a slightly negative impact on prediction performance for heterogeneous graphs that are simplified to homogeneous networks within the pipeline. Each model experienced performance deterioration to a different extent; the largest changes are observed for \textbf{nghb-s}, while the smallest are for \textbf{v-rnk-m}. This is particularly relevant for further research on spreading phenomena due to the evident robustness to perturbations and stability of the \textbf{v-rnk-m} model, also in comparison with methods based on a similar discount-like mechanism. Additionally, it is worth noting the greatest improvement in performance achieved by the \textbf{ts-net} model on homogeneous networks. As for \textbf{Q2}, perceptible changes in the magnitude of the metric emerge only at high reduction ratios ($r \geq0.75$), while results at low-to-moderate~$r$ remain essentially indistinguishable from one another. Together with the structural observations of Sec.~\ref{app:n_flattening}, where \textbf{ForestFire} produces near-identical outcomes across wide ranges of~$r$ and \textbf{TSpanner} moves in the opposite direction to the nominal rate, and with the formal test reported in the previous subsection, this forms a single, coherent picture: the relationship between~$r$ and degradation is non-monotonic, driven by reduction strategy rather than by nominal rate, and only becomes visible once a structural threshold is crossed.

\begin{figure}[ht!]
    \centering
    \includegraphics[width=\linewidth]{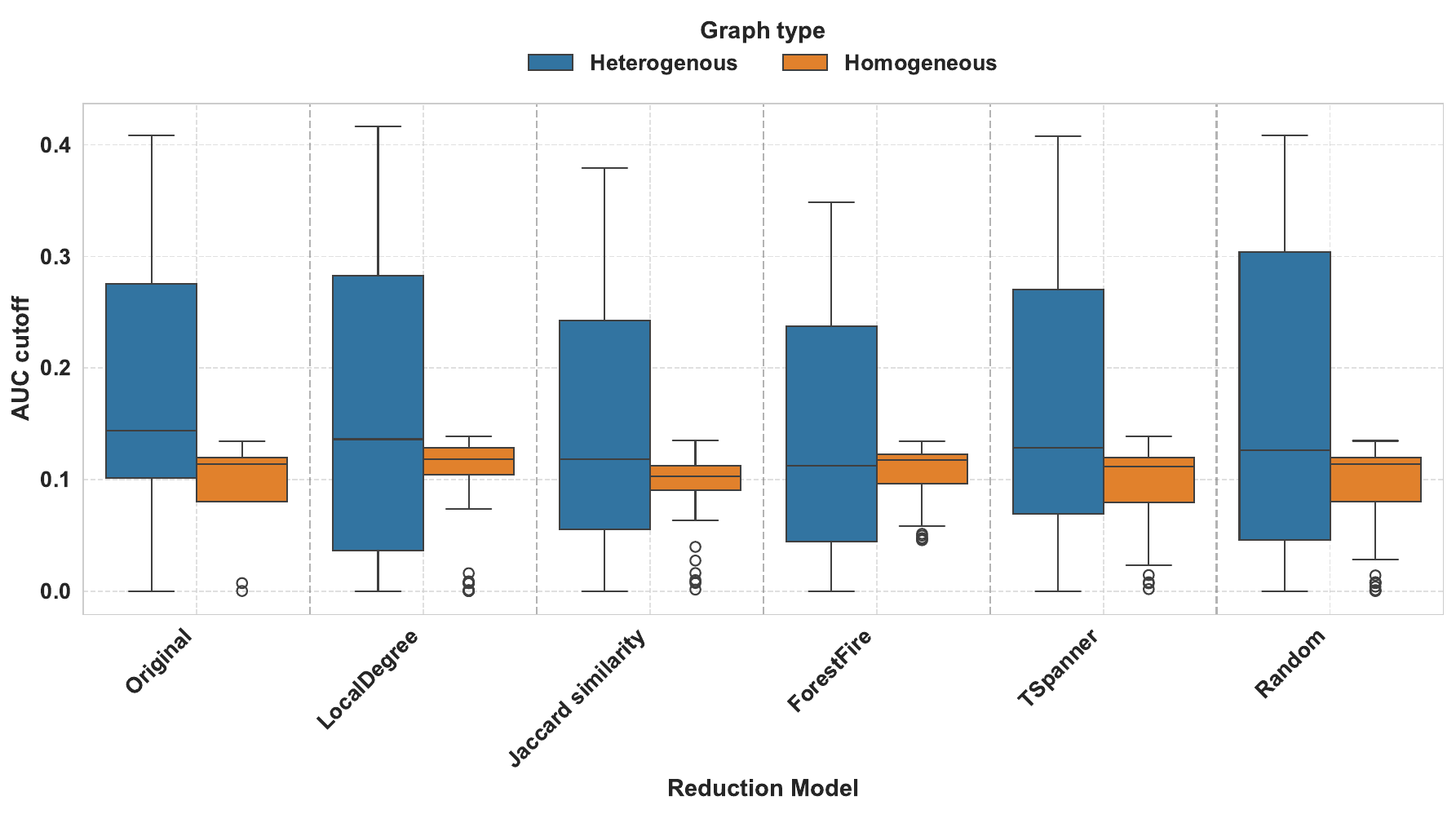}
    \caption{$\mathrm{AUC}_{\mathrm{cutoff}}$ score achieved on the examined networks by seed selection approaches preprocessed using different reduction models.}
    \label{fig:auc_ranking_reduction}
\end{figure}

From another perspective, we may also analyse which sparsification model performs best in the context of ranking prediction, as shown in Fig.~\ref{fig:auc_ranking_reduction}. Although \textbf{TSpanner} achieves the best performance for homogeneous networks in this setting, it attains the lowest maxima for heterogeneous networks. For both types, this model exhibits the most compact distributions within the box plots. In contrast, methods such as \textbf{LocalDegree} or even \textbf{Random} are capable of reaching maximum box-plot values comparable to the original network for heterogeneous graphs; however, their lower quartiles are noticeably reduced.

\section{Conclusion}
This paper introduces a spreading-oriented graph reduction benchmark, \textit{SORB}, which combines $7$ graph reduction algorithms ($5$ sparsification and $2$ coarsening methods) with $6$ influence maximisation models across $7$ networks. Guided by four research questions, our evaluation yields several key findings. With respect to \textbf{Q1} (what is the impact of graph reduction on
the performance of predicted seed sets of different sizes?), sparsification preserves seed set selection quality across all examined seed sizes, with coarsening introducing only marginal performance decline. Regarding \textbf{Q2} (how does the reduction rate r affect downstream results?), downstream degradation is driven primarily by the structural properties of the reduction strategy rather than the nominal reduction rate $r$, with no statistically significant monotonic relationship observed between the two. Addressing \textbf{Q3} (does the application of graph reduction methods lead to consistent effects across diverse input networks?), the two homogeneous networks in our evaluation (\textit{WikiCS}, \textit{Amazon-CS}) exhibit mutually consistent and largely stable behaviour, whereas heterogeneous graphs present a distinct challenge: the flattening transformation required to enable reduction can introduce overhead that counteracts memory savings at lower values of $r$. Finally, with respect to \textbf{Q4} (what is the impact of different graph re- duction methods on the computational resources required by IM methods during prediction?), sparsification yields measurable reductions in both inference time and memory consumption, most notably for the \textbf{v-rnk-m} and \textbf{ts-net} models.

Overall, the results demonstrate significant potential for incorporating sparsification methods as a preprocessing step, reducing memory requirements whilst incurring only small-to-moderate degradation in downstream performance. However, the lack of dedicated approaches for heterogeneous graphs remains a challenge that must be addressed to fully realise this potential. Our platform is open to further extensions in terms of scenarios, data, and models. In future work, we plan to extend it by enabling more effective utilisation of coarsening methods and by introducing support for graph condensation techniques.
With this in mind, we also aim to incorporate methods and processes designed for heterogeneous networks, in order to fully unleash the potential of this preprocessing approach.

\section*{Acknowledgments}
This research was partially supported by the National Science Centre, Poland, grant no. 2022/45/B/ST6/04145, the Polish Ministry of Science and Higher Education programme: International Projects Co-Funded, and the EU under the Horizon Europe, grant no. 101086321. Views and opinions expressed here are those of the authors only and do not necessarily reflect those of the funding agencies.

\section{Appendix A: Environment configuration}
\label{app:training_details}
All experiments were conducted on a machine equipped with an Intel(R) Xeon(R) Gold 6330 CPU @ 2.00GHz, 64~GB RAM, and an NVIDIA A40 GPU. The implementation was developed in Python 3.12.4 using PyTorch 2.5.1. In addition to benchmarking, we introduce a GPU-accelerated PyTorch implementation of the multilayer VoteRank (\textbf{v-rnk-m}) method and released it in version 19.0 of \textit{Network Diffusion}~\cite{czuba2024networkdiffusion}.

\section{Appendix B: Additional results}
\label{app:additional_results}
In this section, we present the remaining results obtained from the conducted experiments. They serve as supplementary material to the results discussed in~\ref{app:experiments}.

\begin{figure}[ht!]
    \centering
    \includegraphics[width=\linewidth]{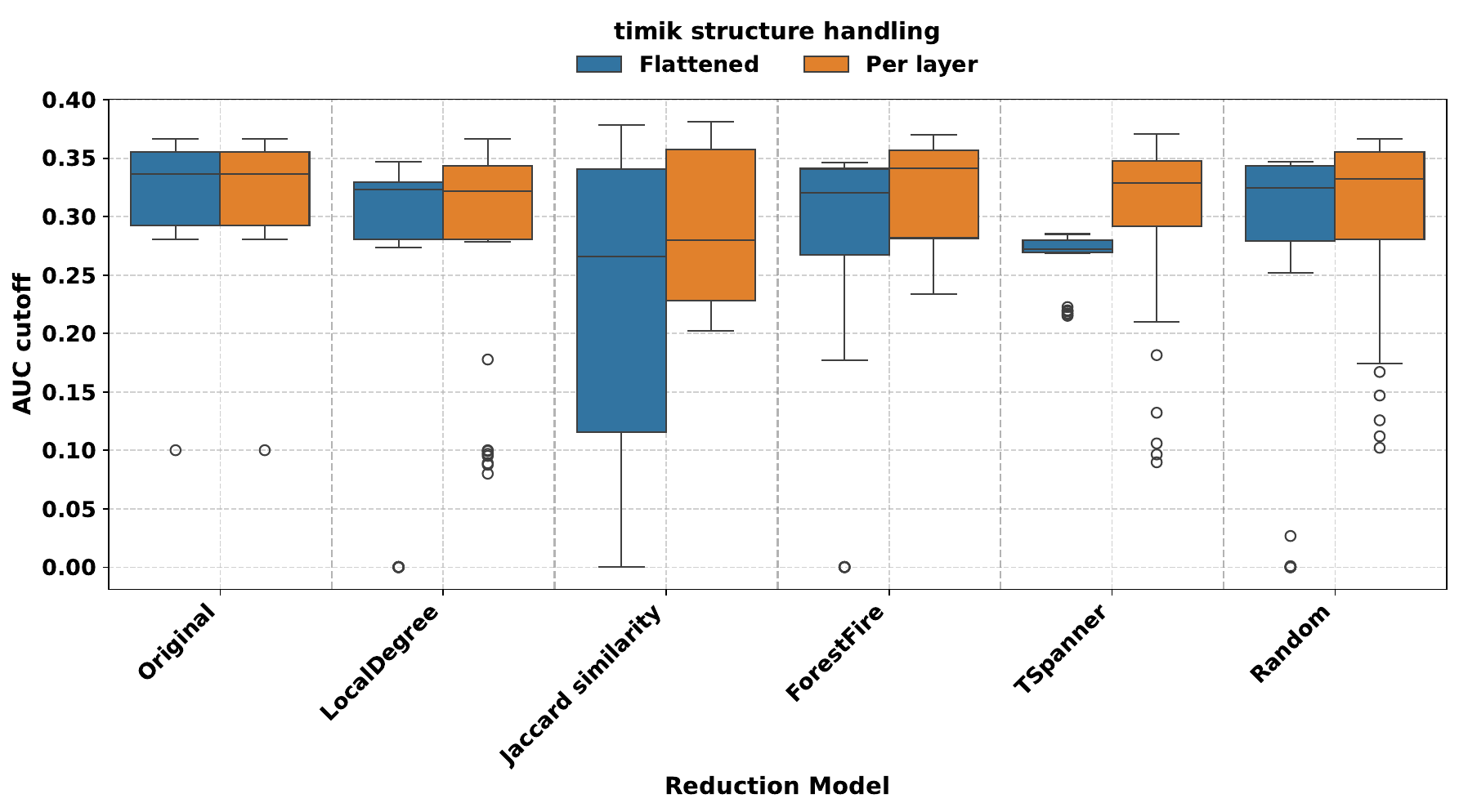}
    \caption{Comparison of $\mathrm{AUC}_{\mathrm{cutoff}}$ on the \textit{timik} network between the two strategies for handling multilayer structure during reduction.}
    \label{fig:timik_per_layer_vs_flattened}
\end{figure}

To verify empirically the natural intuition that multilayer networks should be processed per layer, we compared the $\mathrm{AUC}_{\mathrm{cutoff}}$ of each reduction method on the \textit{timik} network under flattened versus per-layer application, as shown in Fig.~\ref{fig:timik_per_layer_vs_flattened}. We focus the comparison on \textit{timik} as the largest and most complex multilayer dataset in our collection; an analogous analysis on networks with markedly different layer structure (e.g., \textit{FinDKG} with 15 layers, \textit{Freebase} with 3 dense layers) is left for future work. Per-layer processing yields meaningful, though not decisive, improvements, indicating that, at least for this representative case, the observed limitations stem from the reduction methods themselves rather than from the flattened representation they operate on.

In Tab.~\ref{tab:coarsening_auc}, we report the difference between the $\mathrm{AUC}_{\mathrm{cutoff}}$ obtained from predictions on the original graph and on the graph preprocessed with coarsening models. In light of the previously discussed results for these models, the deterioration in performance across every network and for nearly every analysed reduction ratio $r$ is not surprising. Nevertheless, the observations regarding data types are further confirmed. For heterogeneous networks, the decreases are noticeably larger. The performance achieved by the coarsening models may be influenced primarily by our approach, which is based on sorting nodes within the constructed super-nodes according to their degree. In our view, these methods require more in-depth investigation, particularly with respect to processing within super-nodes, which goes beyond the scope of this paper.

\begin{table}[ht!]
    \centering
    \caption{$\Delta\mathrm{AUC}_{\mathrm{cutoff}}$ obtained for the original influence maximisation model results and the score obtained after applying preprocessing with coarsening model.}
    \resizebox{1\linewidth}{!}{%
        \addtolength{\tabcolsep}{0.4em}
        \begin{tabular}{llrrrr}
            \toprule
            reduction model & r & AmazonCS & WikiCS & Freebase & arxiv \\
            \midrule
            \multirow[t]{8}{*}{AffinityGs} & 0.01 & 0.000 & 0.000 & -0.021 & 0.049 \\
             & 0.05 & -0.001 & 0.000 & -0.021 & 0.050 \\
             & 0.10 & -0.003 & -0.002 & -0.022 & 0.049 \\
             & 0.15 & -0.005 & -0.004 & -0.024 & 0.043 \\
             & 0.25 & -0.005 & -0.008 & -0.025 & 0.027 \\
             & 0.50 & -0.027 & -0.019 & -0.049 & -0.082 \\
             & 0.75 & -0.041 & -0.032 & -0.059 & -0.188 \\
             & 0.90 & -0.042 & -0.044 & -0.074 & -0.251 \\
        
            \cline{1-6}
            \multirow[t]{8}{*}{VariationNeighborhoods} & 0.01 & -0.011 & 0.001 & -0.023 & 0.049 \\
             & 0.05 & -0.031 & -0.021 & -0.022 & 0.041 \\
             & 0.10 & -0.047 & -0.030 & -0.023 & 0.033 \\
             & 0.15 & -0.057 & -0.043 & -0.031 & -0.216 \\
             & 0.25 & -0.048 & -0.042 & -0.055 & -0.221 \\
             & 0.50 & -0.052 & -0.035 & -0.076 & -0.249 \\
             & 0.75 & -0.043 & -0.039 & -0.079 & -0.186 \\
             & 0.90 & -0.046 & -0.042 & -0.075 & -0.255 \\
            \bottomrule
        \end{tabular}
    }
    \label{tab:coarsening_auc}
\end{table}

\begin{figure*}[ht!]
    \centering
    \includegraphics[width=\linewidth]{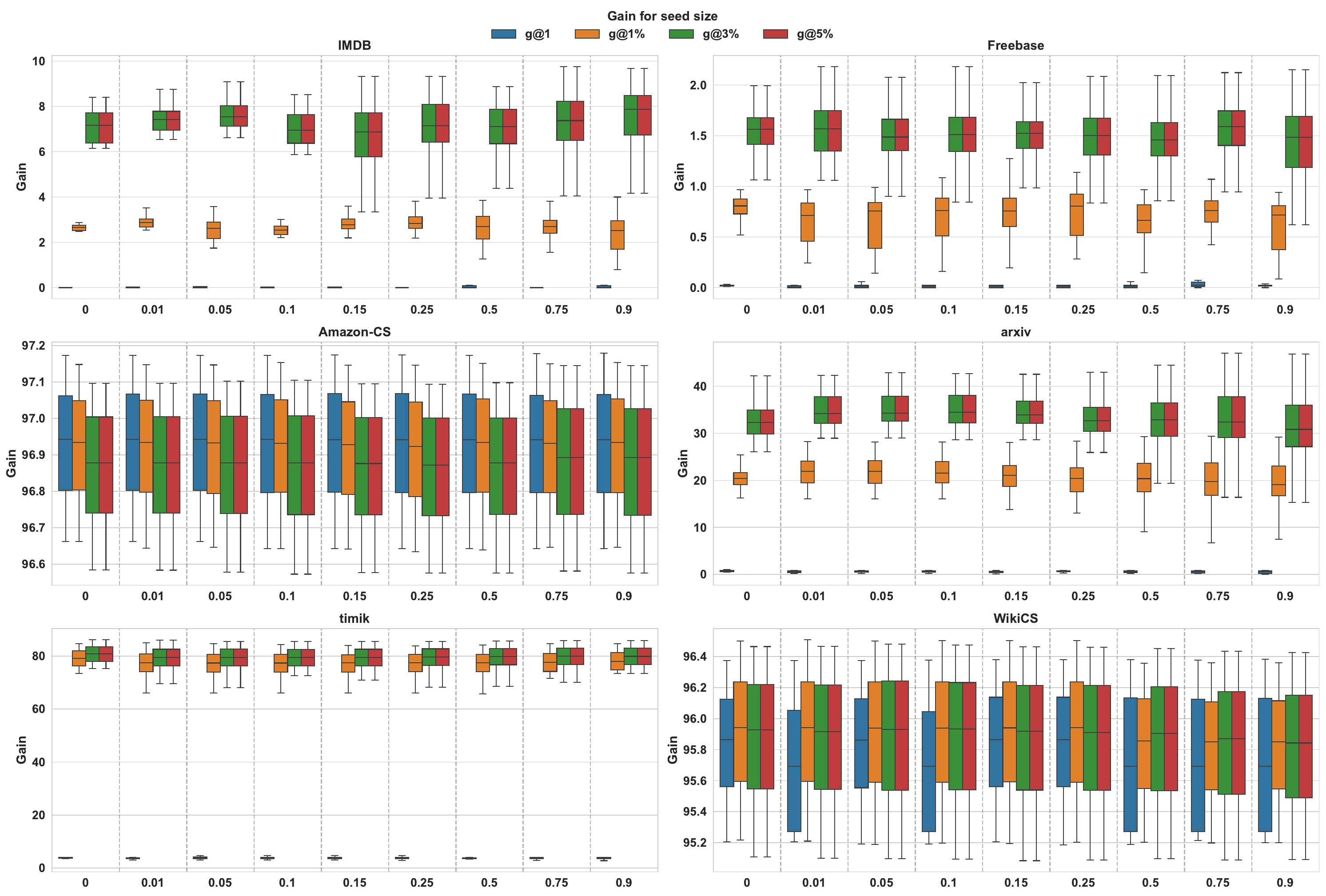}
    \caption{Gain score for each network included in the experiments with the applied sparsification rate $r$.}
    \label{fig:gain_seed_set_rr}
\end{figure*}

\begin{figure}[ht!]
    \centering
    \includegraphics[width=\linewidth]{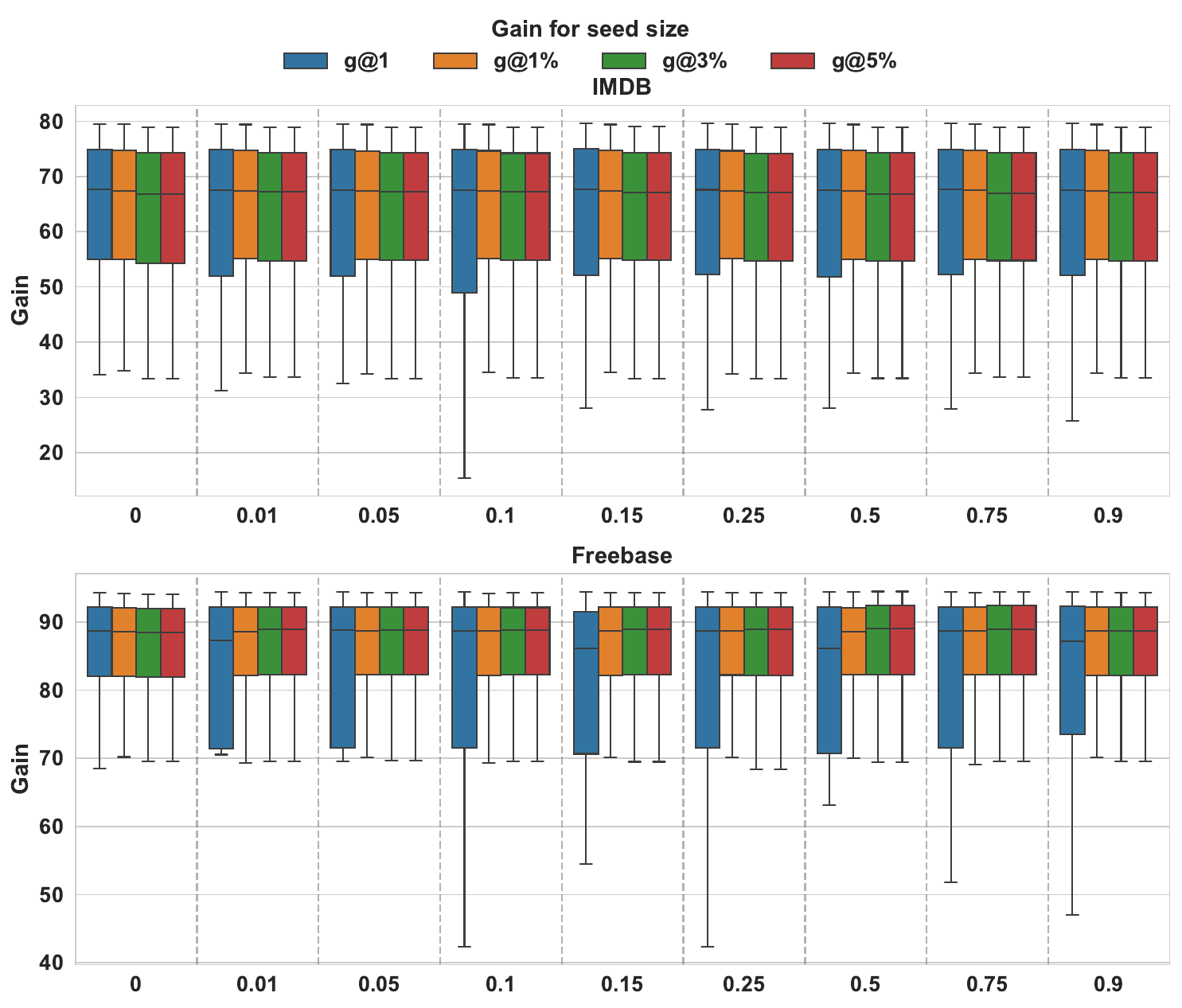}
    \caption{Gain score for film networks with the applied sparsification rate $r$, evaluated using the $OR$ \textbf{MICM} protocol.}
    \label{fig:film_or_gain_seed_set_rr}
\end{figure}

Both coarsening methods respond to~$r$ in the expected direction, with \textbf{AffinityGs} and \textbf{VariationNeighborhoods} exhibiting a clear monotonic deterioration of $\text{AUC}_{\text{cutoff}}$ as~$r$ increases, and with the heterogeneous \textit{Freebase} and \textit{arxiv} networks displaying noticeably larger drops than the homogeneous \textit{AmazonCS} and \textit{WikiCS}. Notably, for \textit{arxiv} both methods even improve over the original baseline at the smallest values of~$r$ before degrading sharply at high reduction rates, suggesting that very mild coarsening can act as a useful denoising step on sparse multilayer graphs whose flattened representation is structurally noisy.

When analysing the individual results for each network presented in Fig.~\ref{fig:gain_seed_set_rr}, the stability of the achieved $Gain$ with respect to the reduction ratio $r$ is evident in most cases. A partial exception is the \textit{arxiv} graph, where a gradual widening of the box plots can be observed, despite the mean often remaining slightly above the baseline. This behaviour may be influenced by certain influence maximisation models that exhibited instability in their prediction results. Additionally, the results obtained for the \textit{IMDB} and \textit{Freebase} networks attracted our attention due to their consistently low $Gain$, regardless of the scenario. This was caused by the highly restrictive requirements of the $AND$ protocol during evaluation with the \textbf{MICM} model and network characteristics. For this reason, we repeated the experiments on these graphs using the $OR$ protocol in the configuration described in~\cite{czuba2025identifying}. Owing to the less restrictive information propagation scheme, the results shown in Fig.~\ref{fig:film_or_gain_seed_set_rr} demonstrate performance levels and stability comparable to those observed for the \textit{timik} and \textit{AmazonCS} networks.

\section{Appendix C: TSpanner adaptation}
\label{app:tspanner_adapt}
A classical $t$-spanner is parameterised by the stretch~$t$; its size follows from~$t$ and the input topology and does not expose a free reduction rate~$r$. To evaluate \textbf{TSpanner} on the same~$r$-axis as the other methods, we use a randomised, budgeted variant with stretch fixed at $t = 4$. Starting from a working copy $G_{\mathrm{c}}$ of the input and an empty output~$G'$ on the same node set, we repeatedly pick a uniformly random edge $(u,v) \in G_{\mathrm{c}}$, remove it from~$G_{\mathrm{c}}$, and add it to~$G'$  if the shortest-path distance between~$u$ and~$v$ in the residual~$G_{\mathrm{c}}$ exceeds~$t$; the loop terminates once $G_{\mathrm{c}}$ has been thinned to $(1-r)\,|E|$~edges. Thus $r$ controls how many edges are subjected to the stretch test rather than a target output size, and~$G'$ contains only those tested edges that proved stretch-critical -- neither the edges remaining in $G_{\mathrm{c}}$ at termination nor the tested but non-critical ones are retained. At small~$r$ few edges are tested and~$G'$ is correspondingly sparse; at large~$r$ almost every edge is tested, the residual graph itself becomes sparser as the loop progresses (so a higher fraction of tests classify edges as critical), and stretch-critical edges accumulate in~$G'$ until its size approaches that of the original graph. The strict $t$-spanner guarantee is sacrificed by the randomised, budget-bounded nature of the test, in exchange for placing \textbf{TSpanner} on the same reduction-rate axis as the remaining methods.

\bibliographystyle{IEEEtran}
\bibliography{references}

\vfill

\end{document}